\documentclass{article}
\usepackage[english]{babel}
\usepackage{float}
\usepackage{booktabs}
\usepackage{graphicx}
\usepackage{dcolumn}
\usepackage{bm}
\usepackage{cite}
\usepackage{amsthm} 
\usepackage[letterpaper,top=2cm,bottom=2cm,left=3cm,right=3cm,marginparwidth=1.75cm]{geometry}
\usepackage{amsmath}
\usepackage{graphicx}
\usepackage[colorlinks=true, allcolors=blue]{hyperref}
\usepackage{authblk}
\usepackage{tabularx}
\usepackage{booktabs}
\usepackage{xcolor}
\usepackage{amsmath}
\usepackage{threeparttable}

\usepackage{longtable}
\usepackage{array}
\usepackage{caption}

\newcolumntype{L}[1]{>{\raggedright\let\newline\\\arraybackslash\hspace{0pt}}m{#1}}
\newcolumntype{C}[1]{>{\centering\let\newline\\\arraybackslash\hspace{0pt}}m{#1}}

\begin{document}
\title{Performances and Correlations of Centrality Measures in Complex Networks}
\author{Yilin Bi, Xinshan Jiao, Tao Zhou}
\affil{CompleX Lab, School of Computer Science and Engineering, University of Electronic Science and Technology of China, Chengdu 610054, China}

\maketitle

\begin{abstract}
Numerous centrality measures have been proposed to evaluate the importance of nodes in networks, yet comparative analyses of these measures remain limited. Based on 80 real-world networks, we conducted an empirical analysis of 16 representative centrality measures. In general, there exists a moderate to high level of correlation between node rankings derived from different measures. We identified two distinct communities: one comprising 4 measures and the other 7 measures. Measures within the same community exhibit exceptionally strong pairwise correlations (all exceeding 0.7 when measured by Kendall’s Tau). In contrast, the remaining five measures display markedly different behaviors, showing weak correlations not only among themselves but also with the other measures. This suggests that each of these five measures likely captures unique properties of node importance. Further analysis reveals that the distribution patterns of the most influential nodes identified by different centrality measures vary significantly: some measures tend to cluster influential nodes closely together, while others disperse them across distant locations within the network. Using the Susceptible-Infected-Recovered epidemic spreading model, we evaluated the performance of those considered measures. We found that LocalRank, Subgraph Centrality, and Katz Centrality perform best in identifying the most influential single node, whereas Leverage Centrality, Collective Influence, and Cycle Ratio excel in identifying the most influential node sets. Overall, measures that identify influential nodes with larger topological distances between them tend to perform better in detecting influential node sets. Interestingly, despite being applied to the same dynamical process, when using two seemingly similar tasks—identifying influential nodes versus identifying influential node sets—to rank the performances of the 16 centrality measures, the resulting rankings are negatively correlated. This further reinforces our conviction that there is no one-size-fits-all centrality measure.
\end{abstract}


\section{Introduction}

Many real-world networks exhibit a heterogeneous structure \cite{Barabasi1999,Holme2005,Caldarelli2007}, meaning that a small fraction of nodes play a critical role in the network’s functioning, while many others are relatively insignificant. A related challenge is how to identify these critical nodes from the network \cite{Pei2013,Lu2016a,Chen2025}. Numerous centrality measures have been proposed to assess the importance of nodes \cite{Freeman1979,Das2018,Rodrigues2018,Wan2021,Bloch2023}. Most measures can be calculated based solely on the network topology \cite{Freeman1977,Bonacich1987,Borgatti2005,Kitsak2010,Morone2015}, while some rely on specific dynamical processes \cite{Dolev2010,Klemm2012,Bauer2012,Liu2016,Ai2022}. The former are more universally applicable, whereas the latter are often suitable only for identifying critical nodes in specific dynamics. This paper primarily focuses on the former.

As hundreds of centrality measures have been proposed \cite{Jalili2015}, a natural question arises: How are these measures interrelated? Analyzing the above question enables us to uncover redundant patterns among centrality measures and assess their distinct informational contributions. If an accurate correlation map is obtained, we can approximate a high-complexity centrality by a strongly correlated low-complexity one, select an appropriate set of centrality measures as node features for machine learning tasks, and characterize the network itself through the interrelationships among centrality measures. In fact, although not very popular, studies in this topic have a long history, dating back to the 1980s \cite{Bolland1988,Rothenberg1995,Faust1997,Estrada2007,Valente2008,Ronqui2015,Li2015,Schoch2017,Oldham2019,Mohamadichamgavi2024,Schoch2024}. Unfortunately, most of the existing studies have one or more of the following three limitations: (i) the centrality measures under consideration are too few, for example, in each of the studies \cite{Bolland1988,Faust1997,Mohamadichamgavi2024,Schoch2024}, no more than four centrality measures were analyzed; (ii) the real networks under consideration are too few, for example, Ref. \cite{Mohamadichamgavi2024} did not analyze any real networks, Ref. \cite{Bolland1988} and Ref. \cite{Rothenberg1995} considered only one network, Ref. \cite{Faust1997} and Ref. \cite{Estrada2007} focused on only one kind of networks, Ref. \cite{Ronqui2015} considered only six networks; (iii) the real networks under consideration are too small, for example, all networks in Refs. \cite{Bolland1988,Faust1997} are of sizes smaller than 100, the average size of networks in each of the studies \cite{Estrada2007,Valente2008,Schoch2017} is smaller than 100. 

Considering the number and diversity of the networks covered, as well as the representativeness and cutting-edge nature of the centrality measures analyzed, Oldham \textit{et al}.'s empirical study \cite{Oldham2019} stands out as the most remarkable. Specifically speaking, they calculated correlations between 17 centrality measures across 212 real networks, showing that the mean correlation for each pair of centrality measures across those 212 networks is generally high, as 97\% of all mean correlations exceed 0.5. Compared to the focus of Oldham \textit{et al}. \cite{Oldham2019}, this work considers additional metrics that were proposed in the past decade and have since attracted widespread recognition. An intriguing finding, previously unreported, is that a few centrality measures exhibit consistently weak correlations with others. For instance, the correlation between \textit{collective influence} \cite{Morone2015} and any other measure is smaller than 0.5, and the average correlations of \textit{leverage centrality} \cite{Joyce2010} and \textit{eccentricity} \cite{Hage1995} with other measures are both below 0.4. We also observe significant differences in the spatial distribution of top-$k$ nodes identified by different centrality measures: under certain measures (e.g., leverage centrality and \textit{cycle ratio} \cite{Fan2021}), the top-$k$ nodes are more dispersed throughout the network, whereas under others (e.g., \textit{eigenvector centrality} \cite{Bonacich1987} and \textit{subgraph centrality} \cite{Estrada2005}), they are closely clustered together. Furthermore, we employ the Susceptible-Infected-Recovered epidemic spreading model \cite{Zhou2006,Pastor-Satorras2015} to evaluate the performance of these measures and find that measures causing top-$k$ nodes to cluster are better at identifying the most influential individual nodes, whereas measures dispersing top-$k$ nodes excel at identifying the most influential node sets. We believe that the results reported in this paper can help future researchers more comprehensively understand the similarities and differences among various centrality measures and better select appropriate measures.

\section{Node Centralities}

Let $G(V,E)$ denote an undirected and unweighted network, where $V$ is the set of nodes and $E$ is the set of edges. Given that hundreds of centrality measures have been proposed to quantify node importance from different perspectives, we select 16 representative ones and provide their detailed definitions below.

\textbf{Degree Centrality} (DC) \cite{Freeman1979} is the simplest and earliest centrality measure that is based solely on the number of immediate neighbors of a node. To ensure comparability across networks and minimize structural biases, a normalization version is typically applied, as 
\begin{equation}
    C^{DC}_{i}=\frac{k_i}{N-1},
\end{equation}
where $k_i$ is the number of neighbors of node $i$, and $N=|V|$ is the number of nodes in the network. $k_i$ is also called \textit{degree} of node $i$.

\textbf{Katz Centrality} (Katz) \cite{Katz1953} extends degree centrality by considering all paths of different lengths, with longer paths weighted less via a damping factor. Its mathematical definition reads
\begin{equation}
    C^{Katz}_{i}=\sum_{k=1}^{\infty} \sum_{j=1}^{N} \alpha^{k} \cdot (\bm{A}^k)_{ij},
\end{equation}
where $\bm{A}$ is the adjacency matrix of the network, $\alpha\in (0,1)$ is the damping factor, and the element $(\bm{A}^k)_{ij}$ denotes the number of paths of length $k$ between node $i$ and node $j$. Usually, $\alpha$ should be chosen sufficiently small to guarantee that the matrix $(I - \alpha \bm{A})$ is invertible, so that the sum in Eq. (2) is convergent.

\textbf{Closeness} \cite{Sabidussi1966} is defined as the inverse of the average shortest-path distance between a node and all other nodes in a connected network, as
\begin{equation}
    C^{Closeness}_i=\frac{N-1}{\sum_{j \neq i} d_{ij}},
\end{equation}
where $d_{ij}$ is the distance between node $i$ and $j$. A node with high closeness is closer to all other nodes, enabling faster information dissemination.

\textbf{Betweenness} \cite{Freeman1977,Newman2001} quantifies the number of shortest paths that pass through a given node, emphasizing its bridging role in the network. It is defined as
\begin{equation}
    C^{Betweenness}_i=\frac{2}{(N-1)(N-2)} \sum_{i \neq s, i \neq t, s \neq t}\frac{g_{st}^{i}}{g_{st}},
\end{equation}
where $g_{st}^{i}$ is the number of shortest paths from $s$ to $t$ passing through $i$, and $g_{st}$ is the total number of shortest paths from $s$ to $t$.

\textbf{Eccentricity} \cite{Hage1995} measures the maximum distance from a node to any other node in the network. Mathematically, it is defined as 
\begin{equation}
    C^{Eccentricity}_i=\frac{1}{\max_{i \neq j} d_{ij}}.
\end{equation}

\textbf{Eigenvector Centrality} (EC) \cite{Bonacich1987} is defined in a self-consistent way, namely a node's score is contributed by all its neighbors, as 
\begin{equation}
    C^{EC}_i=q \sum_{i=1}^{N}a_{ij}C^{EC}_j,
\end{equation}
where $q$ is the proportionality constant. Denote $\overrightarrow{C^{EC}}=\left( C_1^{EC}, C_2^{EC}, \cdots, C_N^{EC} \right)^T$, Eq. (6) can be rewritten in the matrix form as $\overrightarrow{C^{EC}}=q \bm{A} \overrightarrow{C^{EC}}$, which is the eigenvalue equation of $\bm{A}$, with $\overrightarrow{C^{EC}}$ being $\bm{A}$'s eigenvector and $1/q$ being the corresponding eigenvalue. As the dominant eigenvector contains richest information, $q$ is typically set as $q=\frac{1}{\lambda}$, where $\lambda$ being the largest eigenvalue of $\bm{A}$ and thus $C_i^{EC}$ is the $i$-th component of the dominant eigenvector.

\textbf{Information Centrality} (IC) \cite{Stephenson1989} evaluates the influence of nodes within a network through an analogy to electrical circuits. Its mathematical formulation is defined as
\begin{equation}
    C^{IC}_i=\frac{1}{b_{ii}+(T-2R)/N},
\end{equation}
where $T=\sum_{j=1}^{N} b_{jj}$, $R=\sum_{j=1}^{N} b_{ij}$, and $b_{ij}$ denotes the element in the $i$-th row and $j$-th column of the matrix $\bm{B} = (\bm{D}-\bm{A}+\bm{U})^{-1}$. Here, $\bm{D}$ is the diagonal matrix whose diagonal entries are the degrees of the corresponding nodes, and $\bm{U}$ is a matrix with all elements equal to $1$.

\textbf{Subgraph Centrality} (SC) \cite{Estrada2005} quantifies the participation of a node in subgraphs. Estrada and Rodríguez-Velazquez \cite{Estrada2005} only considered a specific kind of subgraphs, say the closed walks (see also the network Hamiltonian defined solely on closed walks \cite{Pan2016}). In a network, the number of closed walks of length $l$ starting and ending on node $i$ is $(\bm{A}^l)_{ii}$, and thus the subgraph centrality of node $i$ is then defined as
\begin{equation}
    C^{SC}_i=\sum_{l=0}^{\infty}\frac{(\bm{A}^l)_{ii}}{l!},
\end{equation}
where $l!$ is utilized to ensure the convergence of the sum. In practical implementation, the subgraph centrality can also be calculated by
\begin{equation}
C^{SC}_i=\sum_{l=1}^{N}(v_l^i)^{2}e^{\lambda_l},
\end{equation}
where $\overrightarrow{v_1}, \overrightarrow{v_2}, \dots, \overrightarrow{v_N}$ is an orthonormal basis composed by eigenvectors of $\bm{A}$ corresponding to the eigenvalues $\lambda_1,\lambda_2, \dots, \lambda_N$, and $v_l^i$ is the $i$-th component of $\overrightarrow{v_l}$.

\textbf{Leverage Centrality} (LC) \cite{Joyce2010} measures the relative advantage of a node by comparing its degree to the degrees of its neighbors. The mathematical formulation is given by
\begin{equation}
    C^{LC}_i=\frac{1}{k_{i}}\sum_{j \in \Gamma_i}\frac{k_{i}-k_{j}}{k_{i}+k_{j}},
\end{equation}
where $\Gamma_i$ denotes the set of neighbors of node $i$.

\textbf{Coreness} \cite{Kitsak2010} (also called $k$-shell index in the literature) records the position of a node in the $k$-core decomposition process \cite{Seidman1983}. The $k$-core decompostion process starts by removing all nodes with degree $k=1$, which causes new nodes with $k\leq 1$ to appear. These nodes are also removed and the process stops when all remaining nodes are of degree $k>1$. The removed nodes and their associated edges form the 1-shell, and any removed node $i$ in 1-shell is assigned a coreness $C_i^{Coreness}=1$. This pruning process is repeated to extract the 2-shell, that is, in each time step the nodes with degree $k\leq 2$ are removed until all remaining nodes are of degree $k>2$. Nodes in the 2-shell are assigned a coreness 2. The provess is continued until all nodes are removed. Recent experiments \cite{Kitsak2010,Morene2019} suggest that coreness is a good measure of a node’s influence in some networked dynamics. 

\textbf{Mixed Degree Decomposition} (MDD) \cite{Zeng2013} takes advantages of both degree and coreness to improve the accuracy of identifying influential nodes. The MDD algorithm defines a so-called mixed degree $k^{(m)}_i$ for each node $i$. Initially, $k^{(m)}_i$ is set to be $i$’s degree $k_i$. The procedure of MDD includes three steps as follows. Step (i): The nodes with minimum mixed degree $M_{\min}$ are removed and added to the $M_{\min}$-shell. Step (ii): Each remaining node $i$'s mixed degree is updated as $k^{(m)}_i\leftarrow k^{(r)}_i+\mu k^{(e)}_i$, where $k^{(r)}_i$ is the number of $i$'s original neighbors in the remaining network, $k^{(e)}_i$ is the number of $i$'s original neighbors having been removed, and $v$ is a tunable parameter in the range [0,1]. If some nodes' updated mixed degrees are no more than $M_{\min}$, they will be removed and added to the $M_{\min}$-shell. Step (ii) will be repeated until all remaining nodes' mixed degrees are larger than $M_{\min}$. All removed nodes constitute the $M_{\min}$-shell and any node $i$ in this shell is assigned a MDD value $C_i^{MDD}=M_{\min}$. Step (iii): Repeat Steps (i) and (ii) until the remaining network is empty. According to the experimental results \cite{Zeng2013}, in this paper, we set $v=0.7$.

\textbf{Lowest Degree Decomposition} (LDD) \cite{Yang2018,Yu2025} identifies vital nodes by sequentially removing the nodes with the lowest degree. Similar to the $k$-core decomposition, in the lowest degree decomposition, the nodes with the lowest degree are firstly removed, which form the 1-shell under LDD and any node $i$ in the 1-shell is assigned an LDD value $C_i^{LDD}=1$. Then, the remaining nodes with the lowest degree are removed, which form the 2-shell and are assigned an LDD value 2. This pruning process stops when all nodes have been removed. A notable difference from $k$-core decomposition is that LDD peels off every shell at once, without any iterations. In fact, it is not hard to rigorously prove that LDD is a subdivision of $k$-core decomposition \cite{Yu2025}. 

\textbf{LocalRank} \cite{Chen2012} evaluates node importance by considering contributions of local paths (up to 4-hop paths), which is defined as 
\begin{equation}
    C^{LocalRank}_i=\sum_{j\in\Gamma_{i}} \sum_{u\in\Gamma_{j}} r(u),
\end{equation}
where $r(u)$ denotes the number of node $u$'s first-order and second-order neighbors.

\textbf{Collective Influence} (CI) \cite{Morone2015} identifies influential nodes based on the optimal percolation. Mathematically terms, it reads
\begin{equation}
    C^{CI}_i=(k_{i}-1)\sum_{j\in\partial Ball(i,l)}(k_{j}-1),
\end{equation}
here, $\partial Ball(i,l)$ denotes the set of boundary nodes of a ball centered at node $i$ with radius $l$. When using CI to rank node influence, the process begins by identifying the node with the highest CI value, which is ranked first. Next, this node is removed from the network, and the CI values of the affected nodes are recalculated. The node with the updated highest CI value is then selected and placed second in the ranking. This iterative process continues, sequentially removing nodes and updating CI values, until the desired number of top-influential nodes is obtained.

\textbf{H-index} \cite{Hirsch2005,Lu2016b} takes into account not only the degree of the target node, but also the degrees of its neighbors. Specifically speaking, H-index of a node $i$, denoted as $C_i^{H-index}$, is defined as the maximum integer satisfying that there are at least $C_i^{H-index}$ neighbors of node $i$ with degrees no less than $C_i^{H-index}$.

\textbf{Cycle Ratio} (CR) \cite{Fan2021} quantifies the proportion of cycles in which a node participates, reflecting its structural stability and feedback potential. As cycles with large sizes are usually less relevant to network functions and to account for all cycles is infeasible due to tremendous computational complexity \cite{Pan2016}, Fan \textit{et al.} \cite{Fan2021} only consider a tiny fraction of cycles with small sizes. Specifically, for each node $i$, among all cycles passing through $i$, they only account for those with smallest size, which constitute the set $S_i$. Denote $S=\cup_{i \in V} S_{i}$ the set of all nodes' shortest cycles, they further define the so-called cycle number matrix $C=[c_{ij}]_{N\times N}$, where $c_{ij}$ is the number of cycles in $S$ that pass through both nodes $i$ and $j$ if $i \neq j$. If $i=j$, $c_{ii}$ is the number of cycles in $S$ that contain node $i$. Then, cycle ratio is defined as
\begin{equation}
    C^{CR}_i=\left\{\begin{array}{c}{{0,c_{i i}=0}}\\ {{\sum_{j,c_{i j}>0}\frac{c_{i j}}{c_{j j}},c_{i i}>0.}}\end{array}\right.
\end{equation}

A comprehensive overview of each centrality measure, including time complexity and related references, is provided in Table \ref{tab:centrality_measures_sorted}.

\begin{table}[htbp]
    \centering
    \caption{\textbf{Computational complexity of the 16 centrality measures}. $N$, $M$ and $\langle k \rangle$ are the number of nodes, the number of edges, and the average degree of the network, with $M \sim N \langle k \rangle$. During the computation of CR, to avoid excessive computational cost caused by nodes being part of very long shortest cycles, a maximum cycle length $L_c$ is typically imposed. If a node $i$ is not contained in any cycle with length not longer than $L_c$, it is approximated that node $i$ is not part of any cycle. The computational complexity of CR thus depends on $L_c$. In the calculation of closeness and betweenness, we assume that the target network is unweighted \cite{Newman2001,Zhou2006a}. The measures are ranked according to their time complexity.}
    \label{tab:centrality_measures_sorted}
    \begin{threeparttable}
        \begin{tabularx}{\textwidth}{>{\hsize=1.5\hsize}X c c >{\hsize=0.5\hsize}X}
            \toprule
            \textbf{Centrality Measures} & \textbf{Time Complexity}& \textbf{References} \\
            \midrule
            Degree Centrality (DC) & $O(M)$ & \cite{Freeman1979} \\
            Leverage Centrality (LC) & $O(M)$ & \cite{Joyce2010} \\
            Coreness & $O(M)$  & \cite{Kitsak2010} \\
            Mixed Degree Decomposition (MDD) & $O(M)$  & \cite{Zeng2013} \\
            Lowest Degree Decomposition (LDD) & $O(M)$  & \cite{Yu2025} \\
            H-index & $O(N \langle k \rangle \log \langle k \rangle)$  & \cite{Lu2016b} \\
            Collective Influence (CI) & $O(N\langle k \rangle ^{l})$  & \cite{Morone2015} \\
            Cycle Ratio (CR) & $O(N \langle k \rangle ^{\frac{L_{c}}{2}})$  & \cite{Fan2021} \\
            LocalRank & $O(N \langle k \rangle^{4})$ & \cite{Chen2012} \\
            Closeness & $O(NM)$  & \cite{Sabidussi1966} \\
            Betweenness & $O(NM)$  & \cite{Freeman1977} \\
            Eccentricity & $O(NM)$  & \cite{Hage1995} \\
            Katz Centrality (Katz) & $O(N^3)$  & \cite{Katz1953} \\
            Eigenvector Centrality (EC) & $O(N^3)$  & \cite{Bonacich1987} \\
            Information Centrality (IC) & $O(N^3)$  & \cite{Stephenson1989} \\
            Subgraph Centrality (SC) & $O(N^3)$  & \cite{Estrada2005} \\
            \bottomrule
        \end{tabularx}
    \end{threeparttable}
\end{table}

\section{Data Description}
This study encompasses an extensive collection of $80$ real-world networks \cite{Rossi2015}, coming from different domains, including biology, sociology, transportation, and so on. Network density spans from $0.0004$ to $0.9908$, reflecting diverse levels of inter-connectivity. A brief summary of the structural properties for these networks is presented in Table \ref{tab:real-networks}. 

\begin{longtable}{L{6cm} C{1.5cm} C{1.5cm} C{1.5cm} C{1.5cm} C{1.5cm}}
\caption{\textbf{Preliminary statistics of the $80$ real-world networks}. $N$ and $M$ denote the number of nodes and edges, $\langle k \rangle$ represents the average degree, $\varepsilon = \frac{2M}{N(N-1)}$ denotes the network density, and $CC$ is the average clustering coefficient.}
\label{tab:real-networks} \\
\hline
\textbf{Network} & \textbf{$N$} & \textbf{$M$} & \textbf{$\langle k \rangle$} & \textbf{$\varepsilon$} & \textbf{$CC$} \\
\hline
\endfirsthead
\caption*{\textbf{Table \thetable} -- continued from previous page} \\
\hline
\textbf{Network} & \textbf{$N$} & \textbf{$M$} & \textbf{$\langle k \rangle$} & \textbf{$\varepsilon$} & \textbf{$CC$} \\
\hline
\endhead
\hline
\multicolumn{6}{r}{\small Continued on next page} \\
\endfoot
\hline
\endlastfoot
aves-songbird-social & 110 & 1027 & 18.6727 & 0.1713 & 0.3274 \\
insecta-ant-colony2-day12 & 111 & 3808 & 68.6126 & 0.6238 & 0.7449 \\
mammalia-voles-bhp-trapping-22 & 115 & 239 & 4.1565 & 0.0365 & 0.3343 \\
insecta-ant-colony3-day30 & 115 & 4321 & 75.1478 & 0.6592 & 0.7704 \\
mammalia-voles-kcs-trapping-25 & 116 & 244 & 4.2069 & 0.0366 & 0.5231 \\
insecta-ant-colony6-day26 & 118 & 4842 & 82.0678 & 0.7014 & 0.7925 \\
ENZYMES g297 & 121 & 149 & 2.4628 & 0.0205 & 0.0512 \\
reptilia-tortoise-network-fi-2012 & 123 & 106 & 1.7236 & 0.0141 & 0.1059 \\
ENZYMES g295 & 123 & 139 & 2.2602 & 0.0185 & 0.0060 \\
ENZYMES g296 & 125 & 141 & 2.2560 & 0.0182 & 0.0059 \\
mammalia-voles-plj-trapping-27 & 125 & 229 & 3.6640 & 0.0295 & 0.4002 \\
aves-wildbird-network-3 & 126 & 1615 & 25.6349 & 0.2051 & 0.7682 \\
mammalia-voles-bhp-trapping-23 & 128 & 253 & 3.9531 & 0.0311 & 0.6053 \\
eco-florida & 128 & 2075 & 32.4219 & 0.2553 & 0.3346 \\
eco-foodweb-baydry & 128 & 2106 & 32.9063 & 0.2591 & 0.3346 \\
aves-wildbird-network-1 & 131 & 1444 & 22.0458 & 0.1696 & 0.8031 \\
insecta-ant-colony2 & 131 & 8437 & 128.8092 & 0.9908 & 0.9919 \\
mammalia-voles-kcs-trapping-27 & 134 & 331 & 4.9403 & 0.0371 & 0.2722 \\
aves-wildbird-network-2 & 135 & 1483 & 21.9704 & 0.1640 & 0.7552 \\
reptilia-tortoise-network-bsv & 136 & 374 & 5.5000 & 0.0407 & 0.1693 \\
insecta-ant-colony3-day14 & 141 & 6607 & 93.7163 & 0.6694 & 0.7660 \\
insecta-ant-colony5 & 152 & 11363 & 149.5132 & 0.9902 & 0.9912 \\
mammalia-dolphin-florida-forage & 190 & 1134 & 11.9368 & 0.0632 & 0.4452 \\
bn-mouse visual-cortex 2 & 193 & 214 & 2.2176 & 0.0116 & 0.0213 \\
aves-wildbird-network & 202 & 4574 & 45.2871 & 0.2253 & 0.8008 \\
bn-mouse brain 1 & 213 & 16089 & 151.0704 & 0.7126 & 0.7583 \\
bn-macaque-rhesus brain 1 & 242 & 3054 & 25.2397 & 0.1047 & 0.4501 \\
econ-wm2 & 257 & 2375 & 18.4825 & 0.0722 & 0.2530 \\
econ-wm3 & 257 & 2379 & 18.5136 & 0.0723 & 0.2653 \\
econ-wm1 & 258 & 2389 & 18.5194 & 0.0721 & 0.2687 \\
inf-USAir97 & 332 & 2126 & 12.8072 & 0.0387 & 0.6252 \\
ca-netscience & 379 & 914 & 4.8232 & 0.0128 & 0.7412 \\
aves-weaver-social & 445 & 1332 & 5.9865 & 0.0135 & 0.6192 \\
bio-celegans & 453 & 2025 & 8.9404 & 0.0198 & 0.6465 \\
bio-celegans-dir & 453 & 2025 & 8.9404 & 0.0198 & 0.6465 \\
power-494-bus & 494 & 586 & 2.3725 & 0.0048 & 0.0420 \\
bio-diseasome & 516 & 1188 & 4.6047 & 0.0089 & 0.6358 \\
fb-pages-food & 620 & 2091 & 6.7452 & 0.0109 & 0.3309 \\
power-662-bus & 662 & 906 & 2.7372 & 0.0041 & 0.0462 \\
power-685-bus & 685 & 1282 & 3.7431 & 0.0055 & 0.1725 \\
rt-twitter-copen & 761 & 1029 & 2.7043 & 0.0036 & 0.0759 \\
soc-wiki-Vote & 889 & 2914 & 6.5557 & 0.0074 & 0.1528 \\
socfb-Reed98 & 962 & 18812 & 39.1102 & 0.0407 & 0.3184 \\
power-1138-bus & 1138 & 1458 & 2.5624 & 0.0023 & 0.0866 \\
inf-euroroad & 1174 & 1417 & 2.4140 & 0.0021 & 0.0007 \\
road-euroroad & 1174 & 1417 & 2.4140 & 0.0021 & 0.0007 \\
power-eris1176 & 1176 & 8688 & 14.7755 & 0.0126 & 0.2163 \\
mammalia-voles-vks-trapping & 1218 & 3592 & 5.8982 & 0.0048 & 0.0967 \\
econ-mahindas & 1258 & 7513 & 11.9444 & 0.0095 & 0.0613 \\
mammalia-voles-plj-trapping & 1263 & 3380 & 5.3523 & 0.0042 & 0.1412 \\
bio-yeast & 1458 & 1948 & 2.6722 & 0.0018 & 0.0708 \\
mammalia-voles-bhp-trapping & 1686 & 4623 & 5.4840 & 0.0033 & 0.1361 \\
power-bcspwr09 & 1723 & 2394 & 2.7789 & 0.0016 & 0.0759 \\
bn-fly-drosophila medulla 1 & 1781 & 8911 & 10.0067 & 0.0056 & 0.0441 \\
tech-routers-rf & 2113 & 6632 & 6.2773 & 0.0030 & 0.2464 \\
rt assad & 2139 & 2786 & 2.6050 & 0.0012 & 0.0143 \\
rt voteonedirection & 2280 & 2464 & 2.1614 & 0.0009 & 0.0019 \\
socfb-nips-ego & 2888 & 2981 & 2.0644 & 0.0007 & 0.0272 \\
inf-openflights & 2939 & 15677 & 10.6683 & 0.0036 & 0.1266 \\
rt damascus & 3052 & 3869 & 2.5354 & 0.0008 & 0.0105 \\
rt obama & 3212 & 3422 & 2.1308 & 0.0007 & 0.0037 \\
rt occupy & 3225 & 3939 & 2.4428 & 0.0008 & 0.0114 \\
rt occupypwallstnyc & 3609 & 3830 & 2.1225 & 0.0006 & 0.0283 \\
rt tlot & 3665 & 4474 & 2.4415 & 0.0007 & 0.0075 \\
rt israel & 3698 & 4164 & 2.2520 & 0.0006 & 0.0037 \\
rt lebanon & 3961 & 4435 & 2.2393 & 0.0006 & 0.0065 \\
rt alwefaq & 4171 & 7059 & 3.3848 & 0.0008 & 0.0865 \\
rt islam & 4497 & 4616 & 2.0529 & 0.0005 & 0.0006 \\
rt bahrain & 4676 & 7977 & 3.4119 & 0.0007 & 0.0170 \\
rt gop & 4687 & 5529 & 2.3593 & 0.0005 & 0.0046 \\
rt p2 & 4902 & 6016 & 2.4545 & 0.0005 & 0.0094 \\
rt oman & 4904 & 6226 & 2.5392 & 0.0005 & 0.0148 \\
inf-power & 4941 & 6594 & 2.6691 & 0.0005 & 0.0801 \\
power-US-Grid & 4941 & 6594 & 2.6691 & 0.0005 & 0.0801 \\
rt libya & 5067 & 5540 & 2.1867 & 0.0004 & 0.0043 \\
ca-Erdos992 & 5094 & 7515 & 2.9505 & 0.0006 & 0.0049 \\
rt uae & 5248 & 6385 & 2.4333 & 0.0005 & 0.0098 \\
power-bcspwr10 & 5300 & 8271 & 3.1211 & 0.0006 & 0.0880 \\
rt dash & 6288 & 7434 & 2.3645 & 0.0004 & 0.0104 \\
\end{longtable}

\section{Correlation Analysis}

We first analyze the correlations between different centrality measures. Given a network $G$ and a centrality measure $C$, based on this measure, we can obtain a score vector $S(G,C)$ that records the centrality values of $G$'s nodes (i.e., $S_i(G,C)$ is the centrality value of the $i$th node). Therefore, for any two measures $C_1$ and $C_2$, we can characterize the correlation between $C_1$ and $C_2$ by calculating the rank correlation between the corresponding score vectors $S(G,C_1)$ and $S(G,C_2)$. We apply the Kendall's $\tau$ \cite{Kendall1938} to quantify the rank correlation. We consider any two vectors associated with all $N$ nodes, $X=(x_1,x_2,\cdots,x_N)$ and $Y=(y_1,y_2,\cdots,y_N)$, as well as the $N$ two-tuples $(x_1,y_1),(x_2,y_2),\cdots,(x_N,y_N)$. Any pair $(x_i,y_i)$ and $(x_j,y_j)$ are concordant if the ranks for both elements agree, namely if both $x_i>x_j$ and $y_i>y_j$ or if both $x_i<x_j$ and $y_i<y_j$. They are discordant if $x_i>x_j$ and $y_i<y_j$ or if $x_i<x_j$ and $y_i>y_j$. If $x_i=x_j$ or $y_i=y_j$, the pair is neither concordant or discordant. Comparing all $N(N-1)/2$ pairs of two-tuples, the Kendall's $\tau$ is defined as
\begin{equation}
	\tau(X,Y)=\frac{2(n_+-n_-)}{N(N-1)},
\end{equation}
where $n_+$ and $n_-$ are the number of concordant and discordant pairs, respectively. If $X$ and $Y$ are independent, $\tau$ should be close to zero, and thus the extent to which $\tau$ exceeds zero indicates the strength of correlation. The values of $\tau$ lies in the range $-1\le\tau\le1$, and the larger value means a stronger correlation. Then, we estimate the correlation between $C_1$ and $C_2$, denoted by $\tau(C_1,C_2)$, as the average Kendall's $\tau$ over all considered networks, say
\begin{equation}
\tau(C_1,C_2)=\frac{1}{N_G}\sum_G \tau(S(G,C_1),S(G,C_2)),
\end{equation}
where $N_G=80$ is the number of considered networks in this study.

\begin{figure}[H]
\centering
\centerline{\includegraphics[width=0.95\linewidth]{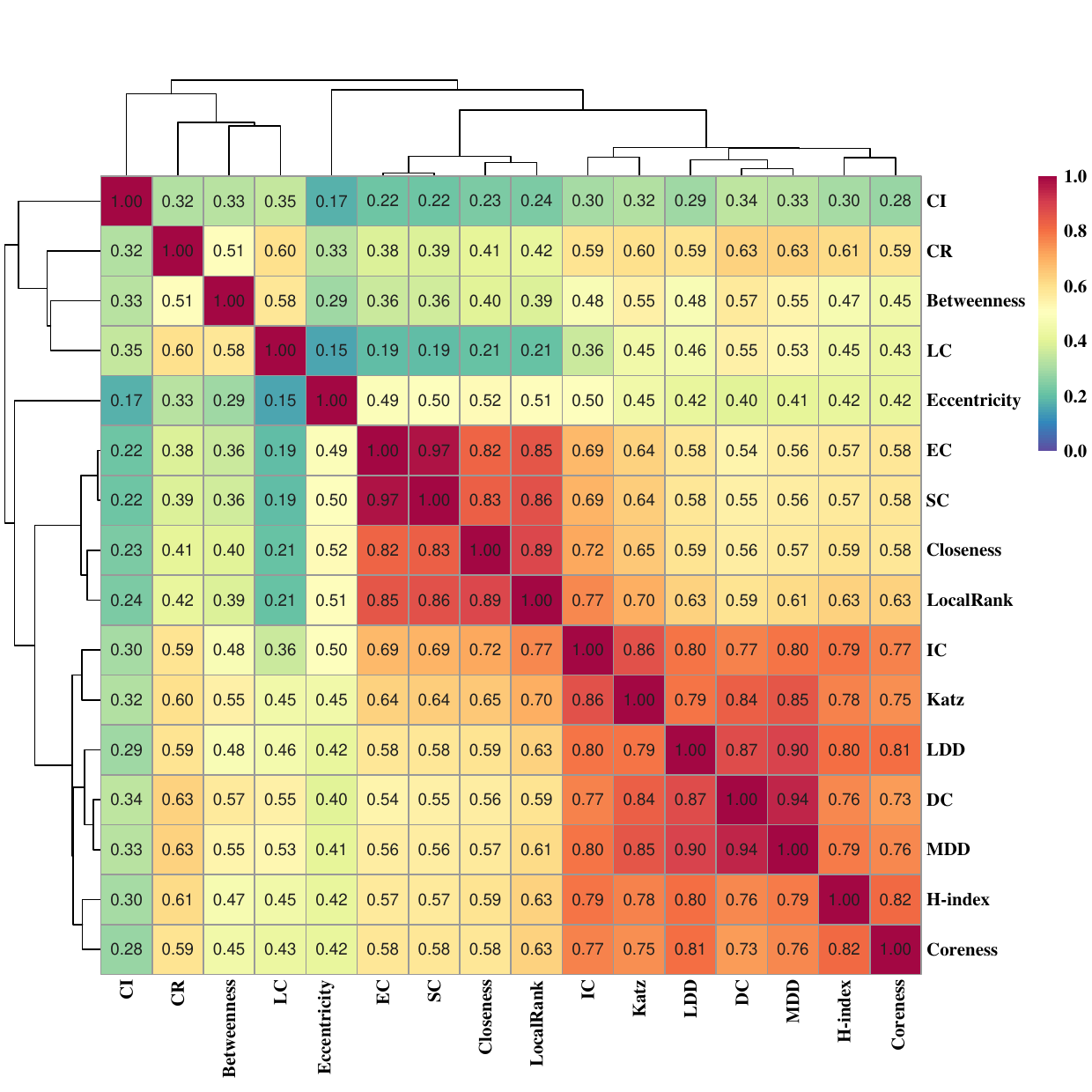}}
\caption{{\bf The average correlations between all pairwise centrality measures over the $80$ real networks.} Different colors indicate different values. The presented hierarchical clustering results are obtained by the algorithm based on average linkage \cite{Seifoddini1986}. }
\label{fig_correlation}
\end{figure}


Figure 1 presents the pairwise correlations among 16 centrality measures. Excluding the trivial case where each measure is perfectly correlated with itself, these 16 centrality measures form a total of 120 pairs. All those 120 correlations are positive, with an average value about 0.55. Overall, these measures exhibit moderate to strong correlations, which is consistent with previous studies. At the same time, in contrast to previous studies—or what they overlooked and failed to report—we found that while all pairwise associations among these centrality measures are positive, their magnitudes vary widely, ranging from a minimum of 0.15 ($\tau(LC,Eccentricity)$) to a maximum of 0.97 ($\tau(EC,SC)$). With the help of a hierarchical clustering algorithm based on average linkage \cite{Seifoddini1986}, we classify the 16 centrality measures into three groups. The first group contains 7 measures (i.e., IC, Katz, LDD, DC, MDD, H-index and Coreness), while the second group contains 4 measures (i.e., EC, SC, Closeness and LocalRank). The pairwise correlations among measures within the same group are all very strong: in the first group, all correlations exceed 0.7, and in the second group, all correlations exceed 0.8. In contrast, the correlations between measures from the first group and those from the second group mostly fall within the range [0.5, 0.7], which are higher than the overall average (0.55) but significantly weaker than the intra-group correlations for the above two groups. The most intriguing is the third group, which includes five centrality measures: CI, CR, Betweenness, LC, and Eccentricity. Their pairwise correlations are relatively low, and their average correlation strengths with other measures are significantly lower than the overall average. We grouped these five measures together not because they form a cluster, but because they all exhibit weaker correlations to other measures. In particular, the correlation between CI and any other measure is smaller than 0.5, and the average correlations of LC and Eccentricity with other measures are both below 0.4.

From Figure 1 and the grouping results, we can derive some valuable insights. First, when dealing with large-scale networks, certain high-complexity measures can be replaced by strongly correlated but low-complexity alternatives. For example, EC and SC, both with a computational complexity of $O(N^3)$, can be replaced by LocalRank, which has a complexity of only $O(N \langle k \rangle^{4})$. Similarly, IC and Katz, also with a complexity of $O(N^3)$, can be replaced by DC, which has a complexity as low as $O(M)$. Second, the measures in the third group exhibit lower correlation strengths with other metrics, and thus are more informative. Therefore, in-depth analysis of these metrics may reveal previously unreported structural properties, and when we need to integrate multiple measures to improve the accuracy of critical nodes identification, these measures may contribute greater incremental value. Although a high information value is an advantage of a measure, we should not simply regard more informative measure as better, because the most important factor in evaluating a measure remains its accuracy. A measure with low accuracy, even if it carries abundant information, holds little practical value. In the next section, we will apply a standard epidemic spreading dynamics to assess the accuracy of each measure in identifying critical nodes under different tasks.

\section{Spreading Influences}
We employ a standard epidemic spreading model, the susceptible-infected-recovered (SIR) model \cite{Zhou2006b,Pastor2015}, to evaluate whether the considered centrality measures can accurately identify the most influential spreaders. This approach is among the most widely used and highly recognized evaluation methods in existing literature. Nodes in a networked SIR model can be in one of three possible states: susceptible, infected and recovered. The SIR process begins with one or more infected seeds and all other nodes are initially susceptible. At each time step, each infected node contacts its neighbors and each susceptible node has an infectivity probability $\beta$ to be infected by one infected neighbor. Then, each previously infected node enters the recovered state with a probability $\gamma$. We set $\gamma=1$ for simplicity. According to the heterogeneous mean-field theory \cite{Castellano2010,Wang2017}, the epidemic threshold of SIR model is approximate to 
\begin{equation}
\beta_c\approx\frac{\langle k\rangle}{\langle k^2\rangle-\langle k\rangle},
\end{equation}
where $\langle k\rangle$ and $\langle k^2\rangle$ denote the mean degree and mean square of degree. 


The first task considered in this study is \textit{the identification of the most influential single node}. For SIR model, the influence of a node $i$, say $I_i$, is defined as the number of eventually recovered nodes averaged over a certain number of independent runs, each of which starts with node $i$ being the sole infected seed. Notice that, when $\beta$ is very small, the disease cannot spread out and the infected node only has a small chance to infect its immediate neighbors, so that the problem to estimate a node’s spreading influence becomes trivial and the best index is just the number of neighbors, say DC. In contrast, when $\beta$ is very high, the disease will infect a large percentage of the population, irrespective of where it originated, and thus the individual influence is meaningless. Accordingly, we focus on the case of $\beta=\beta_c$. Given the target centrality measure $C$ and the corresponding centrality values $C_1, C_2, \cdots, C_N$ of all nodes, we quantify to what extent the measure $C$ resembles spreading influences of individual nodes by two ways. First, we compare the $\rho$ fraction of nodes with the largest $I$ values and the $\rho$ fraction of nodes with the largest $C$ values, and then the {\it Precision} is defined as the proportion of overlap in the total. Second, we directly calculate the rank correlation between centrality values $C_1, C_2, \cdots, C_N$ and node influences $I_1, I_2, \cdots, I_N$, using the Kendall's $\tau$, namely $\tau(C,I)$ (see Eq. 14 and Eq. 15). 

\begin{figure}[H]
\centering
\centerline{\includegraphics[width=0.95\linewidth]{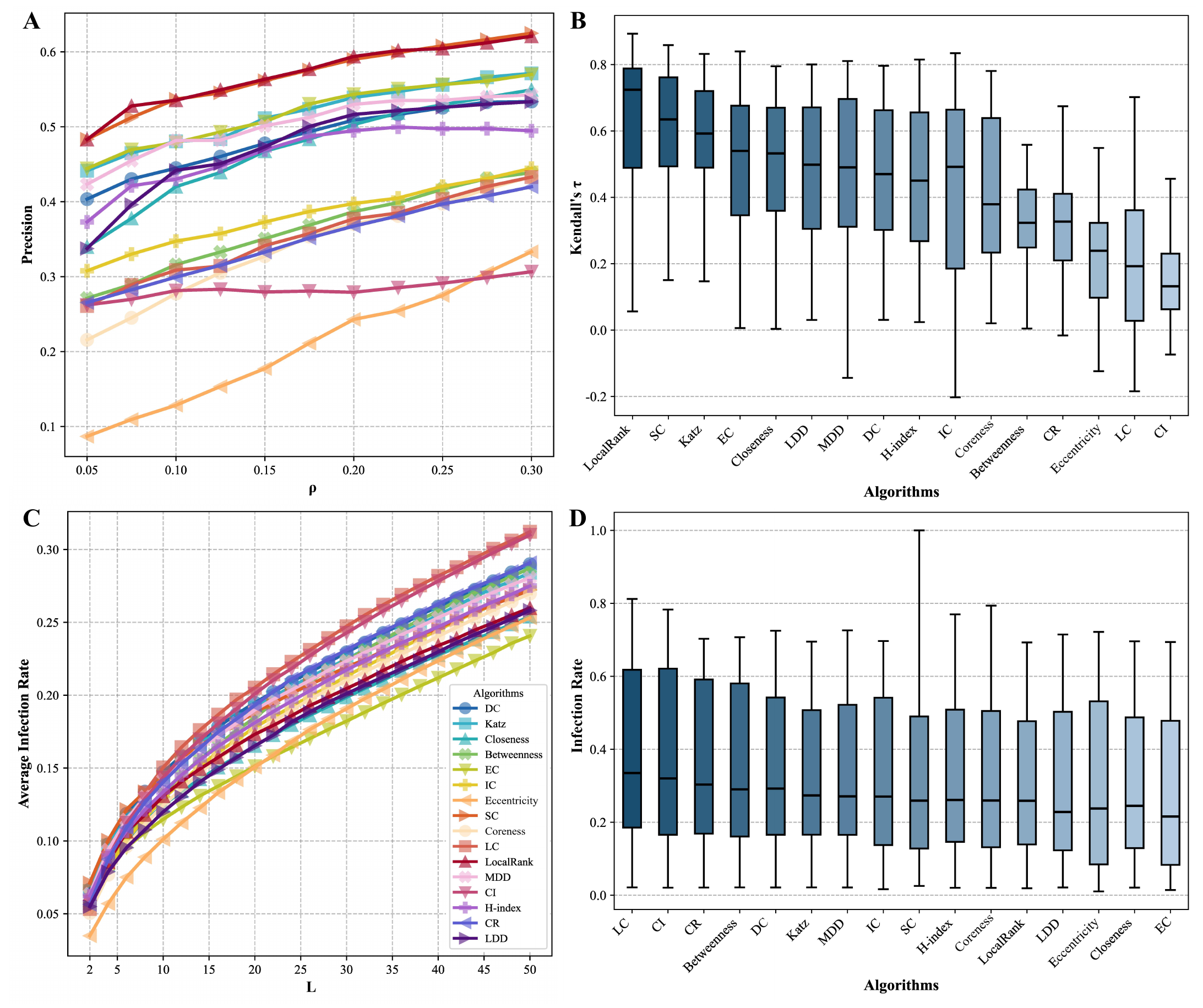}}
\caption{{\bf Performances of the 16 centrality measures under the SIR model.} (A) How Precision changes with increasing $\rho$. (B) The Kendall's $\tau$ between node influences obtained by simulation and node centrality values, ranked in the descending order of the average $\tau$. (C) How the average infection rate changes with increasing $L$. (D) The infection rate for $L=50$, ranked in the descending order of the average values. The data points presented in (A) and (C) represent the average results across 80 networks, with each network's result derived from the average of 10 independent runs. (B) and (D) provide more granular statistics based on the results (Kendall’s $\tau$ and infection rate) for the 80 real networks, with each result again obtained by averaging over 10 independent runs. To enhance visual clarity, for any given algorithm, we did not plot the 80 individual data points corresponding to the 80 real networks. Instead, we adopted the box plot representation, where the horizontal line denotes the median of the 80 data points, and the shaded box covers the interquartile range (IQR), encompassing 50\% of the data points.}
\label{fig_SIR}
\end{figure}

Figure 2A shows how Precision changes with $\rho$. As $\rho$ increases, Precision exhibits an upward trend. This is not surprising, as a centrality measure that randomly assigns scores to nodes would result in Precision equaling $\rho$, which increases linearly with $\rho$. It is worth noting that the Precision of different centrality measures varies significantly, with the top-performing measures being LocalRank and SC, while the worst-performing ones are Eccentricity and CI. Figure 2B presents the Kendall's $\tau$ of those centrality measures, ranked in the descending order of their average $\tau$ over the 80 networks. The results of Figures 2A and 2B are generally consistent. For example, the top-performing measures in Figure 2A, LocalRank and SC, are also ranked in the top two positions in Figure 2B, while the worst-performing measures in Figure 2A, Eccentricity and CI, are ranked third from the bottom and last in Figure 2B, respectively.

The second task considered in this study is \textit{the identification of a set of influential nodes}, which is also named as the \textit{influence maximization problem} \cite{Corley1974,Corley1982,Domingos2001,Richardson2002} in the literature. In the scenario addressed by this study, our task is to identify a set of nodes, given the size $L$ of this set, such that using this set as initial seed nodes maximizes the number of ultimately infected nodes. Accordingly, we use infection rate $r_R$ to quantify the influence of those $L$ nodes, as
\begin{equation}
r_R=\frac{N_R}{N},
\end{equation}
where $N_R$ is the number of recovered nodes after the spreading process ends, which equals the number of ultimately infected nodes as all infected ones will be recovered. Given a centrality measure $C$, in line with most previous studies \cite{Pei2013,Lu2016a,Chen2025}, we employ the greedy algorithm to select this set of nodes. Specifically, we select the $L$ nodes with the highest values from $C_1, C_2, \cdots, C_N$ as this set.  

Figure 2C shows how the average infection rate changes with $L$. Obviously, the average infection rate increases monotonously with the increase of $L$. The performance of different measures exhibits significant variation; however, the overall differences are smaller than those observed in the first task. According to Figure 2C, the two best-performing measures are LC and CI, while the two worst-performing ones are EC and Eccentricity. Figure 2D presents the infection rates of different centrality measures when $L=50$, ranked in the descending order of their average $r_R$ over 80 real networks. The results of Figures 2C and 2D are generally consistent. A phenomenon particularly worth noting is that centrality measures that perform exceptionally well in Figure 2B often underperform in Figure 2D; conversely, those that rank at the bottom in Figure 2B exhibit superior performance in Figure 2D. Indeed, the correlation (also measured by the Kendall's $\tau$) between the two rankings of measures in Figure 2B and Figure 2D is about -0.42. That is to say, when using two similar tasks under the same dynamical process to rank the performances of the 16 centrality measures, the resulting rankings are negatively correlated, indicating that there is no one-size-fits-all centrality measure. 

\section{Topological Analysis}

In the previous section, we observed an interesting phenomenon: measures that perform exceptionally well in the first task often underperform in the second task, and similarly, those that excel in the second task exhibit subpar performance in the first task. To better understand the aforementioned phenomenon, we first examine how the most influential nodes identified by different centrality measures are distributed within the network. We focus on four representative measures, LocalRank, SC, LC, and CI, with the former two being the best-performing measures in the first task and the latter two the best-performing ones in the second task.

Figure 3 presents visualized \textit{rt tlot} network \cite{Rossi2015} with top-1\% nodes produced by each of the above four measures being marked by red. Impressively, the critical nodes selected by LocalRank and SC are clustered in a certain region, while the critical nodes selected by LC and CI are scattered in the whole network. The non-clustered distribution of top-ranked nodes is a significant advantage if one would like to find out a set of critical nodes, because if the selected critical nodes tend to be clustered to each other, their influential areas will be highly overlapped and thus their collective influences are likely to be weak \cite{Zhang2016}.

Since these representative measures exhibit markedly different performance in different tasks (see Figure 2B and Figure 2D), while the distribution patterns of top-ranked nodes produced by them vary significantly in the network (see Figure 3), we propose that the spatial distribution of top-ranked nodes, particularly their dispersion level, serves as a key characteristic of centrality measures. Beyond direct visualization (as Figure 3), we aim to quantify the dispersion level of the top-ranked nodes' distribution. An intuitive approach to achieve this is to calculate the average pairwise distance between these nodes. Provided the top-$L$ nodes $V^t={v^t_1,v^t_2,\cdots,v^t_L}$ by a certain centrality measure, the average distance between top-ranked nodes is defined as 
\begin{equation}
d_t=\frac{2}{L(L-1)}\sum_{1\leq i<j \leq L}d(v^t_i,v^t_j),
\end{equation}
where $d(v^t_i,v^t_j)$ denotes the distance between nodes $v^t_i$ and $v^t_j$. Given a centrality measure, we define $\langle d_t \rangle$ as the average $d_t$ over the 80 real networks and 10 independent runs for each network. 

\begin{figure}[H]
\centering
\centerline{\includegraphics[width=0.95\linewidth]{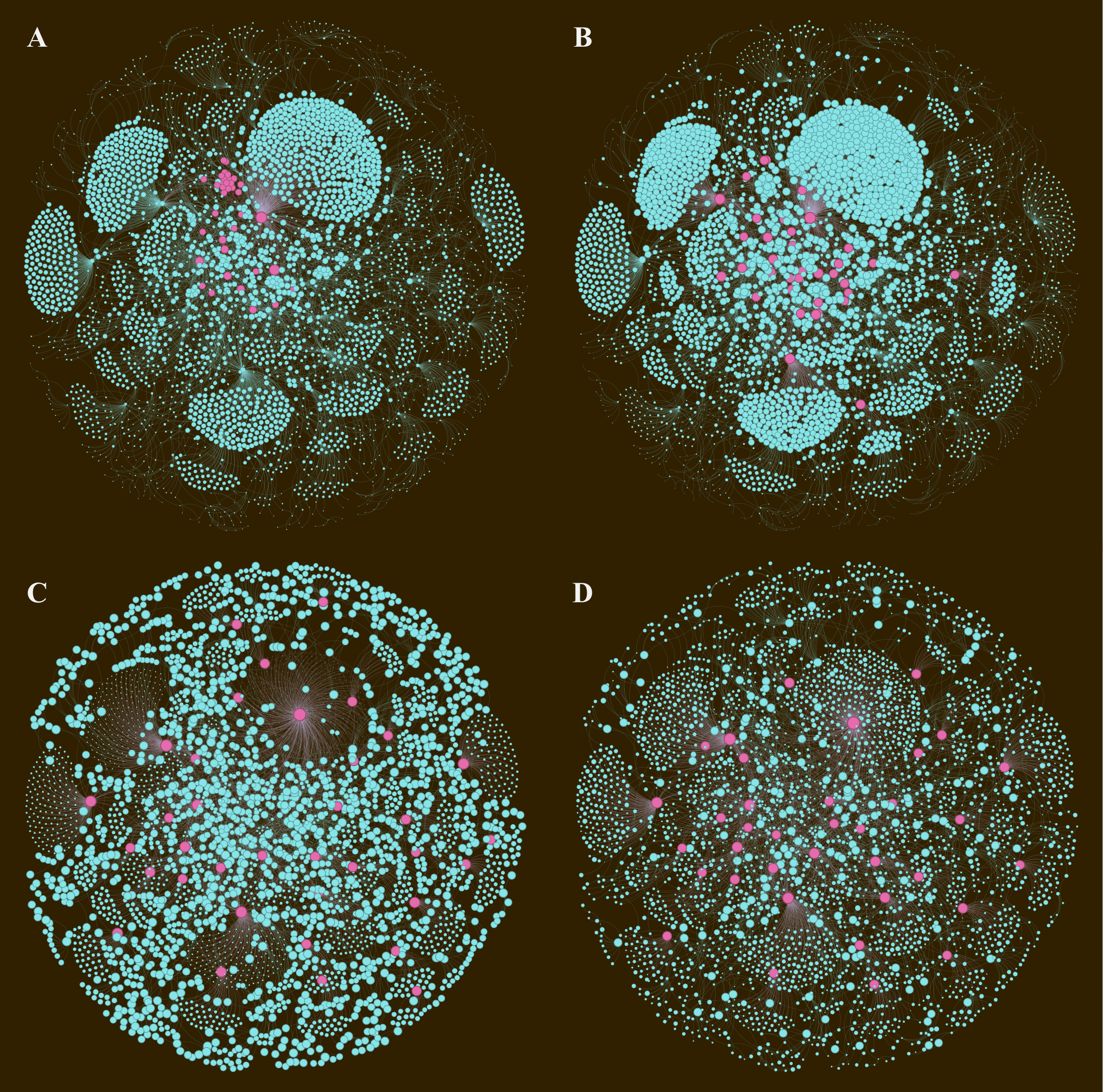}}
\caption{{\bf Visualization of the top-ranked nodes produced by (A) LocalRank, (B) SC, (C) LC, and (D) CI.} The underlying network is \textit{rt tlot}. The top-$1\%$ of nodes (i.e., the top-37 nodes) by the respective centrality measure are highlighted in red. Node size is proportional to its centrality value.}
\label{fig_pca}
\end{figure}

Figure 4 presents how $\langle d_t \rangle$ changes with $L$ under different centrality measures. Based on $\langle d_t \rangle$--$L$ curves, the five measures with the most dispersed top-ranked node distributions (i.e., the highest $\langle d_t \rangle$ values) are LC, CR, Betweenness, Eccentricity, and CI, with LC showing a significantly higher $\langle d_t \rangle$ than the other four. We observe that measures performing well in the second task tend to have high $\langle d_t \rangle$ values. For instance, the top-4 measures in the second task (see Figure 2D) rank within the top 5 in $\langle d_t \rangle$ values, while the 5th-ranked measure in the second task ranks 6th in $\langle d_t \rangle$ values. However, high $\langle d_t \rangle$ alone cannot guarantee strong performance in the second task. For example, Eccentricity has a high $\langle d_t \rangle$ value but ranks third from the bottom in the second task. Therefore, although not strictly, a high $\langle d_t \rangle$ value is approximately a necessary but not sufficient condition for a measure to perform well in identifying a set of influential nodes. A more interesting observation is that the top-5 measures with the highest $\langle d_t \rangle$ values exactly correspond to the 5 measures assigned to the third group based on Figure 1. This is because if a measure yields widely dispersed top-ranked nodes, these nodes are less likely to highly overlap with those from other metrics. Conversely, if two measures both produce clustered top-ranked nodes, their clusters are likely to be highly overlapping, since real-world networks exhibit core-periphery structure \cite{Holme2005,Csermely2013} and rich-club phenomenon \cite{Zhou2004,Colizza2006}, with both clusters potentially belonging to the rich club. In a word, based on the significant relationship between $d_t$ and both the information value of centrality measures and their performance in spreading dynamics, we propose that $d_t$ should be incorporated as a key parameter for characterizing centrality measures.

\begin{figure}[H]
\centering
\centerline{\includegraphics[width=0.95\linewidth]{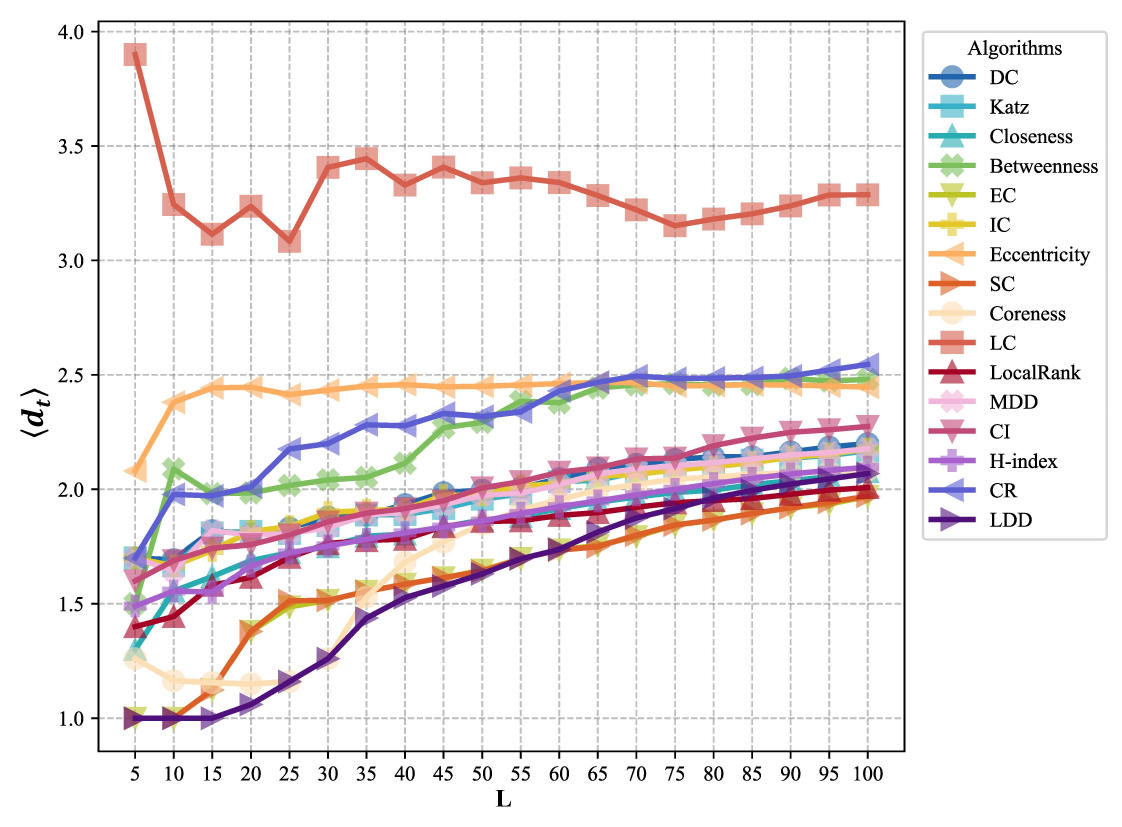}}
\caption{{\bf The average distance between top-ranked nodes produced by the 16 centrality measures.} Each data point represents an average $d_t$ for top-$L$ nodes over 80 real networks and 10 independent runs for each network.}
\label{fig_distance}
\end{figure}

\section{Conclusion and Discussion}

Identifying nodes that play critical roles in a network is of significant theoretical and practical value as well as considerable challenges. Consequently, an increasing number of centrality measures have been proposed to quantify node importance \cite{Pei2013,Lu2016a,Chen2025}. Given this proliferation of diverse measures, analyzing their interrelationships becomes a natural and intuitive question. Indeed, although it has not become a hot topic, sporadic studies on the correlations among centrality measures have appeared in the literature from the 1980s to the present \cite{Bolland1988,Rothenberg1995,Faust1997,Estrada2007,Valente2008,Ronqui2015,Li2015,Schoch2017,Oldham2019,Mohamadichamgavi2024,Schoch2024}. Although the starting point of our analyses is unsurprising, we uncovered intriguing and unreported phenomena during the investigation. First, we identified two clusters where all pairs of measures exhibit very strong correlations. This discovery enables us to substitute computationally expensive measures with simpler alternatives when dealing with large-scale networks. For instance, LocalRank can replace SC ($\tau(LocalRank, SC)=0.86$), as LocalRank’s computational complexity scales linearly with network size, whereas SC’s complexity grows cubically--a critical advantage for sparse large-scale networks. Additionally, we observed some unexpectedly strong correlations between measures lacking obvious mathematical connections. For example, EC and SC show a correlation as high as 0.97, suggesting potential undiscovered mathematical relationships between them. Such observation could provide valuable clues for future studies. Second, we identified five centrality measures that exhibit generally weaker correlations with other measures. This suggests they possess higher information value by capturing node properties not reflected by other measures, warranting special attention. In contrast, previous studies have either overlooked or failed to identify such measures with weak correlations. For instance, in the empirical study by Oldman \textit{et al.} \cite{Oldham2019}, only 3\% of measure pairs showed correlation strength below 0.5, whereas our study reveals over 38\% of such cases. This discrepancy may be attributed to our inclusion of recently proposed measures like CI and CR, which adopt design philosophies distinct from classical measures. The discovery and analysis of these informative measures represent one of the key contributions of this study. 

In addition to the correlation analysis, we further investigate the performances of considered centrality measures by using a well-studied spreading dynamics model, the SIR model. We test those measures by two fundamental tasks: (1) identifying the most influential single node, and (2) identifying the most influential node set. Surprisingly, simulation results revealed that the rankings of measures based on their performances in these two tasks were significantly negatively correlated ($\tau=-0.42$). This indicates that measures performing exceptionally well in the first task often underperform in the second task, and vice versa--those excelling in the second task tend to perform poorly in the first task. In fact, Gu \textit{et al.} \cite{Gu2017} argued that the set of most influential nodes depends strongly on the number of nodes in the set. By a well-designed toy model, They further show that the optimal set of $L$ most important nodes to vaccinate does not need to have any node in common with the optimal set of $L+1$ most important nodes. Although the toy model considered by Gu \textit{et al.} differs significantly from real-world network structures, their work theoretically provides indirect support for the phenomenon we observed: the measures performing exceptionally well in the the first and second tasks can differ. However, our observation goes beyond the mere existence of distinct top-performing measures in the two tasks. It reveals a negative correlation between measure performances across the two tasks within a reasonable range of $L$ values. 

The above phenomenon leads us to consider another research question: how to design a recommender system that generates personalized recommendations satisfying users based on their historical visiting records--in its simplest form, such a system can be conceptualized as link prediction on a bipartite network \cite{Lu2011,Lu2012,Yu2016}. From the user's perspective, a recommender system should simultaneously meet two distinct needs: identifying items they are interested in and discovering items they were previously unaware of. Unfortunately, most recommendation algorithms struggle to fulfill both tasks simultaneously. Indeed, high-accuracy algorithms often have low information value, while algorithms providing novel information are frequently inaccurate. This challenge is known in recommender systems as the \textit{diversity-diversity dilemma} \cite{Ziegler2005,Zhou2010}. Although arising from entirely different contexts, the problem we face in this paper shares similarities with the accuracy-diversity dilemma in recommender systems. While our goal is to identify a single most influential node, the solution simply requires directly targeting the optimal node. However, when the task shifts to identifying a set of most influential nodes, we must avoid overlapping influences, necessitating spatial diversity among selected nodes to maximize coverage across the network. We propose using the average distance $d_t$ among top-ranked nodes to characterize the spatial diversity. Although $d_t$ is defined in a remarkably simple manner, it exhibits strong correlations with both the measure’s performance in identifying a set of most influential nodes and its overall correlation strength with other measures. Therefore, we argue that $d_t$ serves as a key parameter that can reflect the spatial diversity of top-ranked nodes produced by a centrality measure.

In the studies of recommender systems, several methods have been proposed that can achieve both high accuracy and diversity in recommendations by relying solely on network structure information \cite{Lu2012,Yu2016}. One promising approach is to penalize redundant information, thereby enhancing the information value of the algorithm's input and output \cite{Zhou2009}. Motivated by this idea, we believe that measures capable of excelling in both two tasks exist. Such measures must not rely on simple greedy heuristics \cite{Gu2017}; instead, they must penalize nodes near to already selected nodes to avoid redundancy (i.e., to reduce overlapped influence). The strength and extent of the penalty should be strongly dependent on $L$. In conclusion, if we consider the first task as a special case of the second task when $L=1$, to design centrality measures that excel across different values of $L$ remains an open problem that warrants further investigation.





\bibliographystyle{plain}

\end{document}